\documentclass[aps,showpacs,nofootinbib,preprintnumbers,nobalancelastpage,superscriptaddress]{revtex4}

\usepackage[english]{babel}
\usepackage{amsmath,amssymb}
\usepackage[dvips]{graphicx}
\usepackage[sort&compress]{natbib}
\usepackage{amsfonts}
\usepackage{bbm}
\usepackage{color}

\DeclareMathAlphabet{\mathsc}{OT1}{cmr}{m}{sc}

\newcommand{\BR}{\mathrm{BR}}

\allowdisplaybreaks

\def\10{$SO(10)$}
\def\21{SU(2) $\otimes$ U(1) }

\def\422{$SU(4) \otimes SU(2) \otimes SU(2)$}
\def\321{SU(3) $\otimes$ SU(2) $\otimes$ U(1)}

\def\lsim{\raise0.3ex\hbox{$\;<$\kern-0.75em\raise-1.1ex\hbox{$\sim\;$}}}
\def\gsim{\raise0.3ex\hbox{$\;>$\kern-0.75em\raise-1.1ex\hbox{$\sim\;$}}}

\newcommand{\flux}[2][]{\ensuremath{\ifthenelse{\equal{#1}{}}{}{^{#1}\!}\mathit{#2}}}

\newcommand{\AddrAHEP}{%
  AHEP Group, Instituto de F\'{\i}sica Corpuscular --
  C.S.I.C./Universitat de Val{\`e}ncia \\
  Edificio Institutos de Paterna, Apartado Postal 22085, E--46071
  Valencia, Spain}

\newcommand{\fig}[1]{Fig.~\ref{#1}}

\begin{document}

\preprint{IFIC/10-05}


\title{
Probing Neutrino Oscillations in Supersymmetric Models\\
at the Large Hadron Collider}

\author{F.\ de Campos}
\email{camposc@feg.unesp.br}
\affiliation{Departamento de F\'{\i}sica e Qu\'{\i}mica,
             Universidade Estadual Paulista, Guaratinguet\'a, SP,  Brazil }

\author{O.\ J.\ P.\ \'Eboli}
\email{eboli@fma.if.usp.br}
\affiliation{Instituto de F\'{\i}sica,
             Universidade de S\~ao Paulo, S\~ao Paulo, SP, Brazil.}

\author{M.\ Hirsch}
\email{hirsch@ific.uv.es}
\affiliation{\AddrAHEP}

\author{M.\ B.\ Magro}
\email{magro@fma.if.usp.br}
\affiliation{Instituto de F\'{\i}sica,
             Universidade de S\~ao Paulo, S\~ao Paulo -- SP, Brazil.}
\affiliation{Centro Universit\'ario Funda\c{c}\~ao Santo Andr\'e,
             Santo Andr\'e, SP, Brazil.}

\author{W.\ Porod}
\email{porod@physik.uni-wuerzburg.de}
\affiliation{Institut f\"ur Theoretische Physik und Astronomie, 
Universit\"at W\"urzburg, Germany} 
\affiliation{\AddrAHEP}

\author{D.\ Restrepo}
\email{restrepo@uv.es}
\affiliation{Instituto de F\'{\i}sica, Universidad de Antioquia, Colombia}

\author{J.\ W.\ F.\ Valle}
\email{valle@ific.uv.es}
\affiliation{\AddrAHEP}

\begin{abstract}
  The lightest supersymmetric particle may decay with branching ratios
  that correlate with neutrino oscillation parameters.  In this case the
  CERN Large Hadron Collider (LHC) has the potential to probe the
  atmospheric neutrino mixing angle with sensitivity competitive to
  its low-energy determination by underground experiments.  Under
  realistic detection assumptions, we identify the necessary
  conditions for the experiments at CERN's LHC to probe the simplest
  scenario for neutrino masses induced by minimal supergravity with
  bilinear $R$ parity violation.

\end{abstract}


\maketitle

\section{Introduction}
\label{sec:intro}

The CERN Large Hadron Collider (LHC) will provide high enough
center-of-mass energy to probe directly the weak scale and the origin
of mass~\cite{Nath:2010zj,Martin:1997ns,:1997fs,AguilarSaavedra:2005pw,
Ball:2007zza,Aad:2009wy}.
In addition to its designed potential, here we show how LHC searches for
new physics at the TeV region may provide an unexpected opportunity to
probe neutrino properties, currently determined only in neutrino
oscillation experiments~\cite{Schwetz:2008er}, shedding light on some
of the key issues in neutrino physics.
We illustrate how this works in a class of supersymmetric models where the
lepton number is broken, together with the so-called $R$ parity
symmetry~\cite{Barbier:2004ez}. 
Even when the latter holds as a symmetry at the Lagrangian level, as
in some SO(10) unification schemes, $R$ parity breaking may be driven
spontaneously by a nonzero vacuum expectation value of an \321
singlet
sneutrino~\cite{Masiero:1990uj,romao:1992vu,romao:1997xf,Bhattacharyya:2010kr}.
In this case the low-energy theory is no longer described by the
minimal supersymmetric standard model, but contains new
$R$ parity violating
interactions~\cite{hall:1984id,Ross:1984yg,Ellis:1984gi}.
The simplest realization of this scenario leads to an effective model
with bilinear violation of $R$
parity~\cite{Diaz:1998xc,Chun:1998gp,Kaplan:1999ds,
  Takayama:1999pc,Banks:1995by}.
The latter constitutes the minimal way to break $R$ parity in the
minimal supersymmetric standard model
and provides the simplest intrinsically supersymmetric way to induce
neutrino
masses~\cite{Romao:1999up,Hirsch:2000ef,Diaz:2003as,Hirsch:2004he}.
Its main feature is that it relates lightest supersymmetric particle
(LSP) decay properties and neutrino
mixing angles \cite{Porod:2000hv,Choi:1999tq,mukhopadhyaya:1998xj}.
\medskip

Here we demonstrate that indeed, under realistic assumptions, the
simplest scenario for neutrino masses in supersymmetry (SUSY) with bilinear
violation of $R$ parity can be tested at the LHC in a crucial way
and potentially falsified.
We identify the regions of minimal supergravity (mSUGRA) parameters,
event reconstruction efficiencies and luminosities where the LHC will
be able to probe the atmospheric neutrino mixing angle with
sensitivity competitive to its low-energy determination by underground
experiments, both for 7 and 14 TeV center-of-mass energies.

\begin{figure}[!h]
 \centering
\vspace{1pt}
\includegraphics[width=9cm,height=60mm]{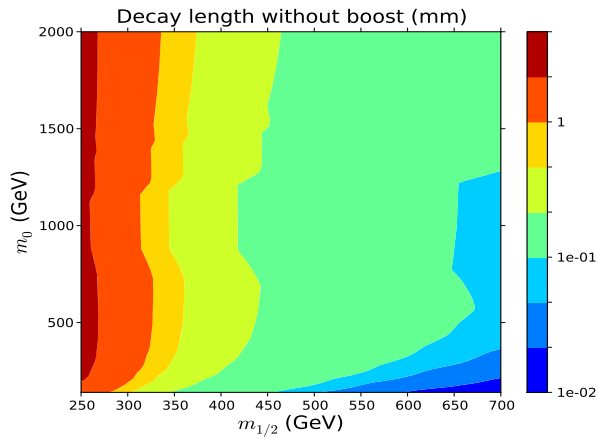}
\caption{$\tilde\chi_1^0$ decay length in the plane $m_0,m_{1/2}$ 
  for $A_0=-100$ GeV, $\tan\beta=10$ and $\mu > 0$.  }
\label{fig:ntrl}
\end{figure}

For the sake of definiteness, we consider the minimal supergravity
model supplemented with bilinear $R$ parity
breaking~\cite{Hirsch:2000ef,Diaz:2003as,Hirsch:2004he} added at the
electroweak scale; we refer to this scenario as RmSUGRA.  In this
effective model one typically finds that the atmospheric scale is
generated at tree level by a \textsl{weak-scale neutralino-exchange
  seesaw}, while the solar scale is induced
radiatively~\cite{Hirsch:2000ef}.
The LSP lacks a symmetry to render
it stable and, given the neutrino mass scales indicated by oscillation
experiments, typically decays inside the LHC
detectors~\cite{Hirsch:2000ef,Porod:2000hv,Diaz:2003as}~\footnote{We
  may add, parenthetically, that such schemes require a different type
  of dark matter particle, such as the axion
  \cite{Chun:1999cq}. Variants with other forms of supersymmetric dark
  matter, such as the gravitino
  \cite{Borgani:1996ag,Takayama:2000uz,Hirsch:2005ag,Staub:2009ww},
  are also possible.}.
As an illustration we depict the neutralino LSP decay length in 
\fig{fig:ntrl}. We can see from \fig{fig:ntrl}
that the expected decay lengths are large enough to be experimentally
resolved, leading to displaced vertex
events~\cite{deCampos:2005ri,deCampos:2007bn}.

More strikingly, one finds that in such a RmSUGRA model one has a strict
correlation between neutralino decay properties measurable at
high-energy collider experiments and neutrino mixing angles determined in
low-energy neutrino oscillation experiments, that is
\begin{equation}
  \label{eq:corr}
  \tan^2\theta_{\mathrm{atm}}\simeq\frac{BR(\tilde\chi_1^0\to \mu^\pm
    W^\mp)}{BR(\tilde\chi_1^0\to \tau^\pm W^\mp)}.
\end{equation}
%
The derivation of Eq.~(\ref{eq:corr}) can be found in \cite{Porod:2000hv}. 
In short, the relation between the neutralino decay branching ratio and 
the low-energy neutrino angle in the bilinear model can be understood 
in the following way. At tree-level in RmSUGRA the neutrino mass matrix 
is given by \cite{Hirsch:2000ef}
\begin{eqnarray}
  \label{eq:NuMass}
m_{eff} &=& 
\frac{M_1 g^2 \!+\! M_2 {g'}^2}{4\, det({\cal M}_{\chi^0})} 
\left(\hskip -2mm \begin{array}{ccc}
\Lambda_e^2 
\hskip -1pt&\hskip -1pt
\Lambda_e \Lambda_\mu
\hskip -1pt&\hskip -1pt
\Lambda_e \Lambda_\tau \\
\Lambda_e \Lambda_\mu 
\hskip -1pt&\hskip -1pt
\Lambda_\mu^2
\hskip -1pt&\hskip -1pt
\Lambda_\mu \Lambda_\tau \\
\Lambda_e \Lambda_\tau 
\hskip -1pt&\hskip -1pt 
\Lambda_\mu \Lambda_\tau 
\hskip -1pt&\hskip -1pt
\Lambda_\tau^2
\end{array}\hskip -3mm \right) 
\end{eqnarray}
where $\Lambda_i = \mu v_i + v_D \epsilon_i $ and $\epsilon_i$ and 
$v_i$ are the bilinear superpotential parameters and scalar neutrino 
vacuum expectation value, respectively. Equation~(\ref{eq:NuMass}) is 
diagonalized by two angles; 
the relevant one for this discussion is the angle 
$\tan\theta_{23} = - \frac{\Lambda_{\mu}}{\Lambda_{\tau}}$. One can 
understand this tree-level mass as a seesaw-type neutrino mass with 
the right-handed neutrino and the Yukawa couplings of the ordinary 
seesaw replaced by the neutralinos of the minimal supersymmetric
standard model and couplings of the 
form $c \Lambda_{i}$, where $c$ is some combination of (generation 
independent) parameters. These couplings, which determine 
(the generation structure of) the neutrino mass matrix, also 
determine the couplings $\chi_i^0-l_i^{\pm}-W^{\mp}$ and 
$\chi_i^{\pm}-\nu_i-W^{\mp}$ \cite{Porod:2000hv}. Taking the 
ratio of decays to different generations the prefactors $c$ drop 
out and one finds Eq.~(\ref{eq:corr}), when the angle 
$\tan\theta_{23}$ is identified with the atmospheric neutrino 
angle. One-loop corrections tend to modify this relation, but, 
as long as the loop corrections are smaller than the tree-level 
neutrino mass, Eq.~(\ref{eq:corr}) is a good approximation 
\cite{Porod:2000hv}. 

\begin{figure}[!h]
 \centering
\vspace{1pt}
\includegraphics[width=80mm,height=60mm]{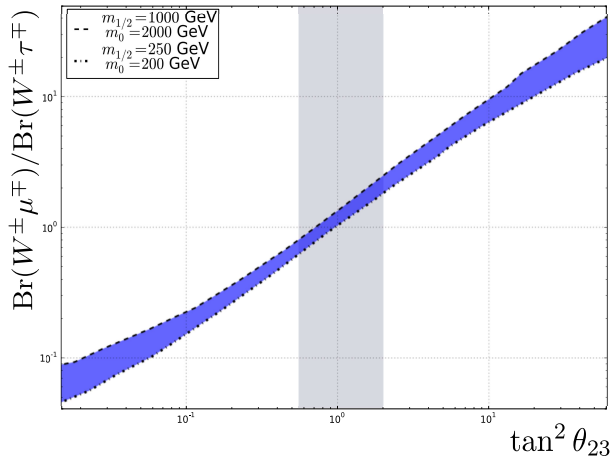}
\caption{Ratio of $\tilde\chi_1^0$ decay branching ratios, Br$(\tilde
  \chi^0_1\to \mu q'{\bar q})$ over Br$(\tilde \chi^0_1\to \tau
  q'{\bar q})$ in terms of the atmospheric angle in bilinear $R$ parity
  violation~\cite{Porod:2000hv}. The
  shaded bands include the variation of the model parameters in such a
  way that the neutrino masses and mixing angles fit the required
  values within $3\sigma$.}
\label{fig:ntrl2}
\end{figure}

In other words, as seen in \fig{fig:ntrl2}, the LSP
decay pattern is predicted by the low-energy measurement of the
atmospheric angle~\cite{Porod:2000hv,Romao:1999up}, currently
determined by underground low-energy neutrino
experiments~\cite{Schwetz:2008er}, as
$$
 \sin^2\theta_{\mathrm{atm}} = 0.50^{+0.07}_{-0.06}
$$
the 2 and 3~$\sigma$ ranges being 0.39--0.63 and 0.36--0.67,
respectively.

In this paper we show how a high-energy measurement of LSP decay
branching ratios at the LHC allows for a redetermination of
$\theta_{\mathrm{atm}}$ and hence a clear test of the model.  We provide
quantitative estimates of how well this ratio of branchings should be
measured at LHC in order to be competitive with current oscillation
measurements.  This issue has already been addressed but only at the
parton level, using some semirealistic acceptance and reconstruction
cuts, and for just one specific mSUGRA point~\cite{Porod:2004rb}.

\section{Framework of our analysis}
\label{sec:cuts}

Our goal is to present a more detailed analysis of the LHC potential
to measure the LSP branching ratios required to test the relation
shown in Eq.~(\ref{eq:corr}), going beyond the approximations made in
the previous work of Ref.~\cite{Porod:2004rb}. The generation of the
supersymmetric spectrum and decays in the scope of the RmSUGRA model
was carried out using the \texttt{SPheno}
package~\cite{porod:2003um}\footnote{An updated version including
  bilinear $R$ parity violation can be obtained at
  \texttt{http://www.physik.uni-wuerzburg.de/$\sim$porod/SPheno.html}.}.
The event generation was done employing
\texttt{PYTHIA}~\cite{sjostrand:2006za} with the RmSUGRA particle
properties being passed into it in the SUSY Les Houches accord (SLHA)
format~\cite{skands:2003cj,Allanach:2008qq}. Jets were defined using
the subroutine \texttt{PYCELL} with a cone size of $\Delta R
=0.4$. \medskip

A striking property of RmSUGRA models is the existence of displaced
vertices associated to the LSP decay \cite{deCampos:2007bn}. We use 
the detached vertices to
probe the LSP branching ratio relation Eq.~(\ref{eq:corr}). In order
to mimic the LHC potential to study displaced vertices we use a toy
detector based on the ATLAS technical
proposal~\cite{:1997fs}. \medskip

We begin our analysis demanding that the events pass some basic
requirements to guarantee that they will be triggered by the
experimental collaborations.  This is done because the LHC experiments
have not defined so far any specific strategy to trigger displaced
vertices with such high invariant mass, therefore, we restricted our
analysis to events that would be accepted by the ongoing analyses.  We
accept events passing at least one of the following requirements,
denoted as cut {\bf C1},
\begin{enumerate}

\item the event has one isolated electron or a photon with $p_T> 20$ GeV;

\item the event has one isolated muon with  $p_T> 6$ GeV;

\item the event has two isolated electrons or photons with $p_T> 15$ GeV;

\item the event has one jet with $p_T> 100$ GeV;

\item the event has missing transversal energy in excess of $100$ GeV.

\end{enumerate}

Next, in cut {\bf C2}, we require that at least one of the neutralinos in the
event decays beyond the primary vertex point, that is, outside an ellipsoid
~\cite{deCampos:2007bn}
\begin{equation}
  \label{eq:minellip}
      \left ( \frac{x}{5\delta_{xy}} \right )^2
   +  \left ( \frac{y}{5\delta_{xy}} \right )^2
   +  \left ( \frac{z}{5\delta_{z}} \right )^2   = 1 \; ,
\end{equation}
where the $z$ axis is taken along the beam direction. We made a
conservative assumption, since we are not performing a detailed
detector simulation, that the ellipsoid dimensions are 5 times the
ATLAS expected resolutions in the transverse plane ($\delta_{xy} =
20~\mu$m) and in the beam direction ($\delta_z = 500~\mu$m), in order
to ensure that the neutralino displaced vertex is distant of the
primary vertex.  We also demand that all tracks must be initiated
inside the pixel inner detector within a radius of $550$ mm and
$z$ axis length of $800$ mm. A detached vertex complying with these
requirements we called \emph{signal vertex}. \medskip

In order to check relation Eq.~(\ref{eq:corr}) we looked for detached
vertices presenting a $W$ associated to them and we must isolate the
LSP decays into $W\mu$ and $W\tau$.  Moreover we consider only
hadronic final states of the $W$ as a necessary condition for the
identification of the lepton flavor. In cut {\bf C3}, which is
designed for the
$W$ reconstruction, we require two jets with charged tracks
intersecting the neutralino resolution ellipsoid, and invariant mass
between 60 and 100~GeV. In order to be sure that the W reconstruction
is clean, we further impose that the axes of other jets of the event
to be outside of a cone $\Delta R=0.8$ of the $W$ jets' axes.  Note
that this cut should eliminate standard model (SM) backgrounds coming
from displaced
vertices associated to $b$'s or $\tau$'s.  To guarantee a high quality
in the reconstruction of the displaced vertices we impose that the $W$
decay jets must be central, having pseudorapidities $| \eta| < 2.5$;
this constitutes our cut {\bf C4}.  The events passing the above
requirements most probably originate from LSP decay, having basically
no sizable standard model background, except for instrumental
backgrounds and beam-gas interactions.  \medskip

A signal vertex is classified as originating from the LSP decay into a
$\mu W$ pair if it presents a $\mu^\pm$ and a hadronically decaying
$W$ stemming from the displaced vertex with transverse momentum $p_T>
6$ GeV and $| \eta| < 2.5$.  In the $\tau^\pm$ case we demanded that
the $\tau^\pm$ associated to a detached $W$ possesses $p_T> 20$ GeV
and $|\eta|<2.5$. These requirements are called {\bf C5}.  \medskip

Detecting taus is somewhat more complicated than detecting muons, so
one needs to be more careful in reconstructing the $\tau W$ pair
displaced vertex.  The following criteria, denoted {\bf C6}, are used
to separate the detached vertices exhibiting a $\tau^\pm$ through its
1- and 3-prong decay modes. We check also that the secondary displaced
vertex from tau decay does not spoil the signal vertex; {\em i.e.},  we
verify that the tau decay products point towards the LSP decay vertex
within the experimental resolution. We define the neutralino
resolution ellipsoid as the ellipsoid centered at the displaced vertex
position of neutralino, $v_1$, with axes $\delta_{xy} = 12~\mu$m and
$\delta_z = 77~\mu$m based on ref.~\cite{:1997fs}.  Let
$\mathbf{p}_{\text{prong}}$ be the momentum of either 1--prong tau decay
or the sum of momenta of the 3--prong decays. Let also $v_2$ be the
position of the secondary vertex coming from $\tau$. We verify whether
the line along $\mathbf{p}_{\mathrm{prong}}$, crossing $v_2$
intersects the neutralino resolution ellipsoid. For this we require
that for each $\tau$, the discriminant of quadratic equation for
parameter $t$
\begin{equation}
  \label{eq:intsph}
  \sum_i^2
  \left(
\frac{p_{\text{prong}}^it+v_2^i-v_1^i}{\delta_{xy}}
  \right)^2+
  \left(
\frac{p_{\text{prong}}^3t+v_2^3-v_1^3}{\delta_z}
  \right)^2-1=0
\end{equation}
be equal to or greater than zero.  In previous \cite{Porod:2004rb}
analysis only 3-prong tau decays modes were considered. \medskip

An additional cut {\bf C7} was applied to 3--prong tau events, {\em
  i.e.}  we also require that one of the prongs has a transverse
momentum $p_T > 9$ GeV while the other two have $p_T > 2$ GeV. In
addition we check if all prongs lie within a cone radius of $\Delta R
< 0.2$ around the tau direction obtained from the prongs' tracks.
\medskip

Finally we require that the signal lepton ($\mu$ or $\tau$) be
isolated; cut {\bf C8}. $\mu$ isolation demands that there are no
other tracks whose total transverse energy satisfies $E_T>5$~GeV within
a cone $\Delta R>0.3$. The $\tau$ was required to be isolated using
the same criteria as for the muon, but for an annulus of outer radius
$\Delta R=0.4$ and inner radius $\Delta R=0.1$.  Isolation of the
leptons is a needed requirement to eliminate events presenting leptons
generated inside jets and constitutes an important cut to reduce
potential backgrounds.\medskip

\section{Results and discussion}
\label{sec:res}

In order to access the effects of the above defined cuts {\bf
  C1}--{\bf C8} we present detailed information on their effects for
the mSUGRA SPS1a benchmark point \cite{Allanach:2002nj} characterized
by $m_{1/2}=250\,$GeV, $m_{0}=100\,$GeV, $A_0=-100\,$GeV,
$\tan\beta=10$, and $\mathrm{sgn}(\mu)=+1$. This allows us to compare
our results with the one previously obtained in
\cite{Porod:2004rb}. For the default solution of \texttt{SPheno} to the
neutrino masses and mixings, the relevant neutralino branching ratios
are
\begin{align}
  &\BR(\tilde\chi_1^0\to W^\pm\mu^\mp)=5.4\%
   &&\BR(\tilde\chi_1^0\to W^\pm\tau^\mp)=6.2\%
    &&\BR(\tilde\chi_1^0\to Z\nu)=1.2\%\nonumber
\\
  &\BR(\tilde\chi_1^0\to e^\pm\tau^\mp\nu)=11.5\%
   &&\BR(\tilde\chi_1^0\to\mu^\pm\tau^\mp\nu)=24.3\%
    &&\BR(\tilde\chi_1^0\to\tau^\pm\tau^\mp\nu)=36.4\%\nonumber
  \label{eq:br}
\\
  &&& \BR(\tilde\chi_1^0\to b\bar{b}\nu)=14.7\%; 
\end{align}
with the $R$ parity parameters being 
\begin{align}
  &\epsilon_1=0.0405~\mathrm{GeV},
   &&\epsilon_2=-0.0590~\mathrm{GeV},
    &&\epsilon_3=0.0506~\mathrm{GeV},\nonumber
\\
  &v_1=-0.0027~\mathrm{GeV},
   &&v_2=0.0042~\mathrm{GeV},
    &&v_3=-0.0033~\mathrm{GeV}.\nonumber
  \label{par:sps1} 
\end{align}
Furthermore, for this choice of parameters the neutralino decay length
is $ \label{eq:ct} c\tau=1.1\,\mathrm{mm}, $ and it travels an average 
of 4.4~mm in the laboratory.

\begin{table}
  \begin{tabular}{|c|r|r|r|r|r|}\hline
    cut    &$N_\mu$  &$N_\tau$
&$N_{\tau\to\mathrm{all}}^{\mathrm{1-prong}}$&$N_{\tau\to\mathrm{hadron}}^{\mathrm{1-prong}}$&$N_\tau^{\mathrm{3-prong}}$\\\hline
{\bf C1} & 0.996 &  0.968 &  0.816 &   0.475  &  0.058 \\ 
{\bf C2} &  0.923 &  0.898 &   0.757 &   0.440  &  0.055 \\ 
{\bf C3} &  0.391 &  0.407 &   0.344 &   0.199  &  0.025 \\ 
{\bf C4} &  0.369 &  0.385 &   0.325 &   0.188  &  0.024 \\ 
{\bf C5} &  0.230 &  0.248 &   0.211 &   0.121  &  0.024 \\ 
{\bf C6+C7} &  0.230 &  0.078 &  0.057 &  0.033  &  0.014 \\
{\bf C8} &  0.102 &   0.015 &   0.014 &   0.009  &   0.001 \\ \hline
  \end{tabular}
  \caption{Fraction of events passing the successive cuts {\bf C1--C8}
    used for the event reconstruction at the SPS1a mSUGRA point. }
\label{tab:cuts}
\end{table}

\begin{figure}
  \centering
\includegraphics[scale=0.4]{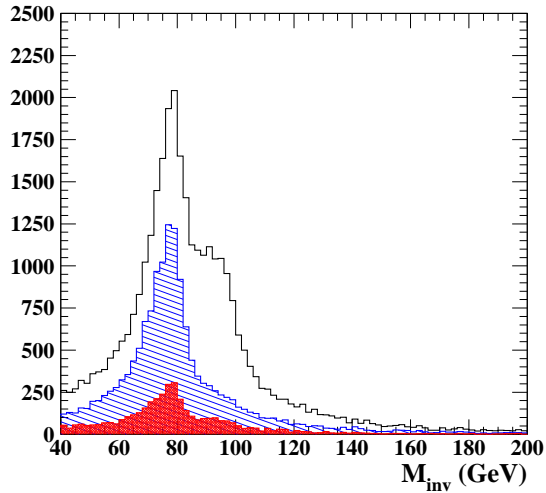}
  \caption{ From top to bottom: $\tilde\chi_1^0 \to j j X$ without
    cuts, with cut on lepton isolation ($\mu$ or $\tau$) and with all
    other cuts leaving free the invariant mass range. }
  \label{fig:mwinv}
\end{figure}

From Table \ref{tab:cuts} we see that the vast majority of the
events pass the trigger requirements {\bf C1}, as expected. For the
SPS1a SUSY point, the LSP decay length is sufficiently long to
guarantee that a sizeable fraction of its decays take place away from
the primary vertex; this reflects as a high efficiency for passing the
cut {\bf C2}. We have focused our attention to events presenting a
$W^\pm$ decaying into two jets through {\bf C3}.  It is interesting to
notice that 63\% of the $W$ hadronic decays are in the form of two
jets. Additional suppression of the signal by {\bf C3} comes from the
matching of the sum of momenta of the charged tracks pointing to the
detached vertex and the jets reconstructed using \texttt{PYTHIA}.  To further
illustrate the $W$ decay, we present in \fig{fig:mwinv} the
jet--jet invariant mass distribution.  As we can see, this
distribution is clearly peaked around the $W$ mass and a good fraction
of the two jets reconstructed as associated to the LSP decay  pass
the cut {\bf C3}. The observed high efficiency of cut {\bf C4} shows
that the $W$'s produced in the LSP decay are rather central. \medskip

We also learn from Table~\ref{tab:cuts} that  detached vertices
presenting a $W$ possess  around 60\% of the time
an energetic $\mu^\pm$ or $\tau^\pm$
complying with {\bf C5}.  Moreover the cuts
{\bf C6} and {\bf C7}, which ensure the quality of the $\tau$
reconstruction, reduce significantly the number of $W^\pm \tau^\mp$
events. Finally the isolation cut {\bf C8} turns out to be quite
important significantly reducing the signal.  \medskip

For the parameter point SPS1a, the expected efficiencies for the
reconstruction of $\mu W$ and $\tau W$ decays are 0.107 and 0.0098 
respectively, where in the last we have
added 1-- and 3--prong hadronic decays. When the $\tau$ decays into
a $\mu$ and neutrinos, the event was computed as being a $\mu W$ decay
if the $\mu$ passes the cuts. This was included appropriately in our
calculations. Taking into account the total SUSY production cross
section (41 pb) at 14 TeV, an integrated luminosity of 100 fb$^{-1}$ and these
efficiencies we anticipate that the number of observed $\mu W$ and
$\tau W$ events after cuts to be
\begin{align}
  N_\mu=&32000 & N_\tau^{\text{hadron}}= 3382
\nonumber
\end{align}  
where
$N_\tau^{\text{hadron}}=N_{\tau\to\mathrm{hadron}}^{\mathrm{1-prong}}+N_\tau^{\mathrm{3-prong}}$.
Therefore, the statistical accuracy of the ratio
$R=BR(\tilde\chi_1^0\to \mu^\pm W^\mp)/BR(\tilde\chi_1^0\to \tau^\pm
W^\mp)$ is expected to be $\sigma(R)/R=\sqrt{1/N_\mu +1/N_\tau }\approx
0.015$. In the case one takes into account only the three-prong decays
of the tau, as in Ref.\ \cite{Porod:2004rb}, the statistical error 
of this ratio increases to $\approx 0.053$. 
Moreover, as expected, there is a degradation of the
accuracy in the determination of this ratio of branching ratios in a
more realistic analysis; the result obtained in \cite{Porod:2004rb} is
$\simeq 0.028$.  \medskip

In the  evaluation of the above efficiencies we have not taken into
account multiple interactions at the LHC as needed for the high
luminosity run. Therefore, we reevaluated the detection efficiencies
for muons and taus with multiple interactions switched on in
\texttt{PYTHIA}. We found that these efficiencies were only slightly
degraded by the occurrence of pileup, that is, we obtained that the
efficiencies for muon reconstruction are reduced to $0.102$ and
for tau are $0.008\;68$ in hadronic mode and $0.000\;94$ in
the 3-prong mode.  In our analyses we took into account the effect of
multiple interactions. \medskip

For the sake of comparison, we present a detailed analysis for a
different mSUGRA point that is $m_{1/2}=500\,$GeV, $m_{0}=500\,$GeV,
$A_0=-100\,$GeV, $\tan\beta=10$, and $\mathrm{sgn}(\mu)=+1$.  Once
again using \texttt{SPheno}, we obtain that the neutralino branching ratios
larger than $1\%$ are:
\begin{align}
 &\BR(\tilde\chi_1^0\to W^\pm\mu^\mp)=22.9\% , 
  &&\BR(\tilde\chi_1^0\to W^\pm\tau^\mp)=25.2\% ,
   &&\BR(\tilde\chi_1^0\to Z\nu)=25.1\%,
\nonumber
\\
&\BR(\tilde\chi_1^0\to \nu h^0)=16.9\% ,
  &&\BR(\tilde\chi_1^0\to\tau^\pm\tau^\mp\nu)=3.4\% ,
    &&\BR(\tilde\chi_1^0\to b\bar{b}\nu)=2.9\%;
\nonumber
\end{align}
and the corresponding $R$ parity parameters are 
\begin{align}
  &\epsilon_1=0.1507~\mathrm{GeV},
   &&\epsilon_2=-0.1507~\mathrm{GeV},
    &&\epsilon_3=0.1507~\mathrm{GeV},\nonumber
\\
  &v_1=-0.0056~\mathrm{GeV},
   &&v_2=0.0058~\mathrm{GeV},
    &&v_3=-0.0054~\mathrm{GeV}.\nonumber
\end{align}
As we can see, the neutralino LSP decays are dominated by the
two--body ones, in contrast with the SPS1a point where the three--body
decays mediated by light scalars are dominant. Because of its heavier
spectrum, the total SUSY production for this parameter point is
smaller than the SPS1a one; however, the cross section loss is
partially compensated by the higher branching ratios into $\mu W$ and
$\tau W$. 

The total cross section for this case is $832.0$ fb and our analyses
indicate that the reconstruction efficiency for $\mu W$ decays is
0.203 while the $\tau W$ decays are reconstructed with an efficiency
of 0.035, where we did not take into account pileup.  The inclusion
of this effect leads to a tiny reduction of the reconstruction
efficiencies that become 0.199 for $\mu W$ and 0.033 for $\tau W$. On
the other hand the efficiency for reconstructing a $\tau W$ event in
the 3--prong mode is 0.012.  Notice that these efficiencies are larger
for this mSUGRA point than for the SPS1a because the neutralino is
heavier and, consequently, its decay products are more energetic and
pass the cuts more easily.  The expected total number of reconstructed
events after cuts for this SUSY point is $ N_\mu=5171$ and $
N_\tau^{\text{hadron}}=933$ where we have included the pileup
effects.  Therefore, the expected statistical error on the ratio $R$
becomes $\approx0.036$, or $\approx0.056$ when we only use 3--prong
taus as in \cite{Porod:2004rb}.  As we can see, the statistical error
on the ratio $R$ increases as $m_{1/2}$ (LSP mass) increases due to
the reduction of the SUSY production cross section despite the
increase in the detection efficiencies. \medskip

%
\begin{figure}
  \centering
  \includegraphics[height=70mm,scale=0.4]{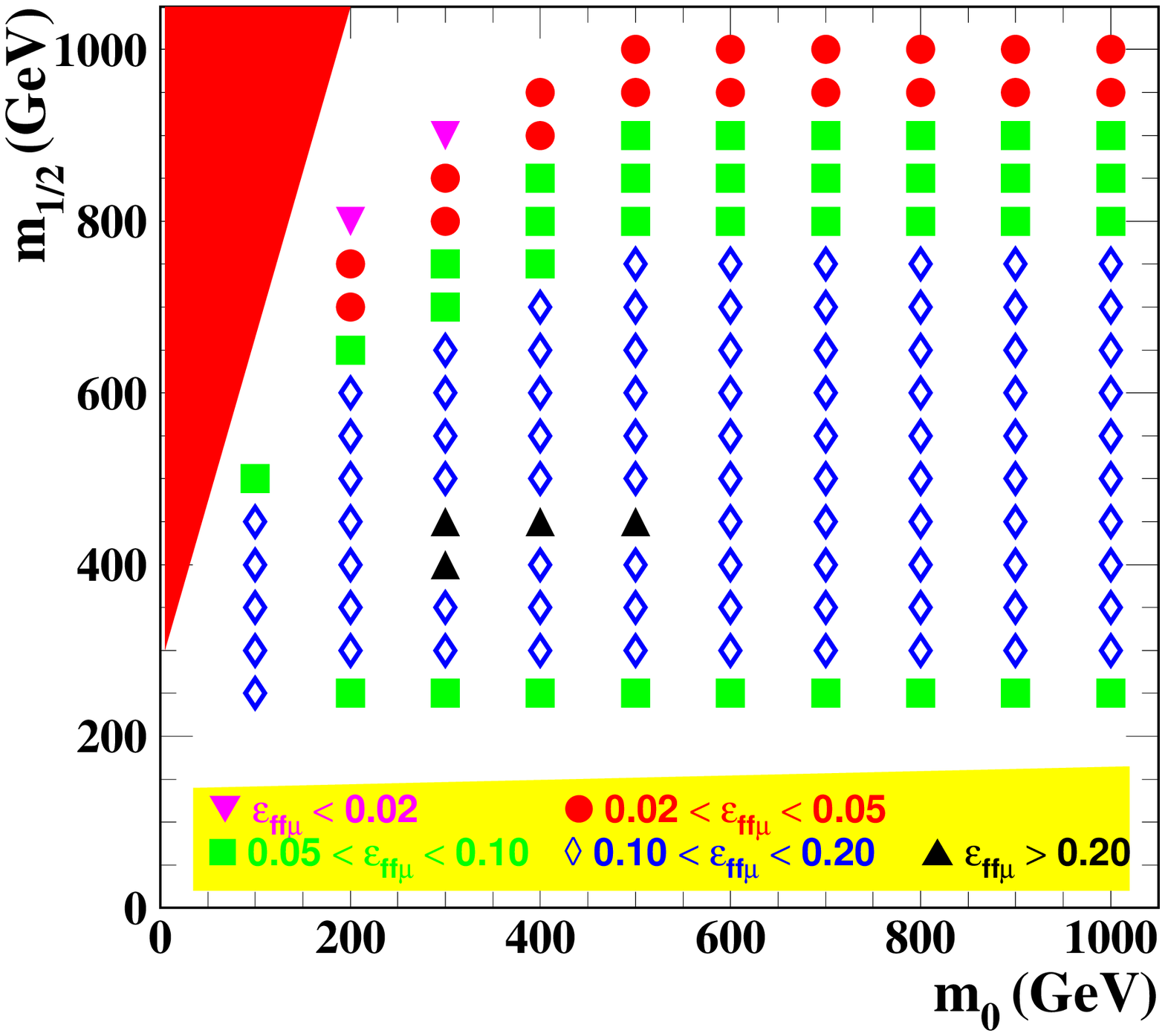}
  \includegraphics[height=70mm,scale=0.4]{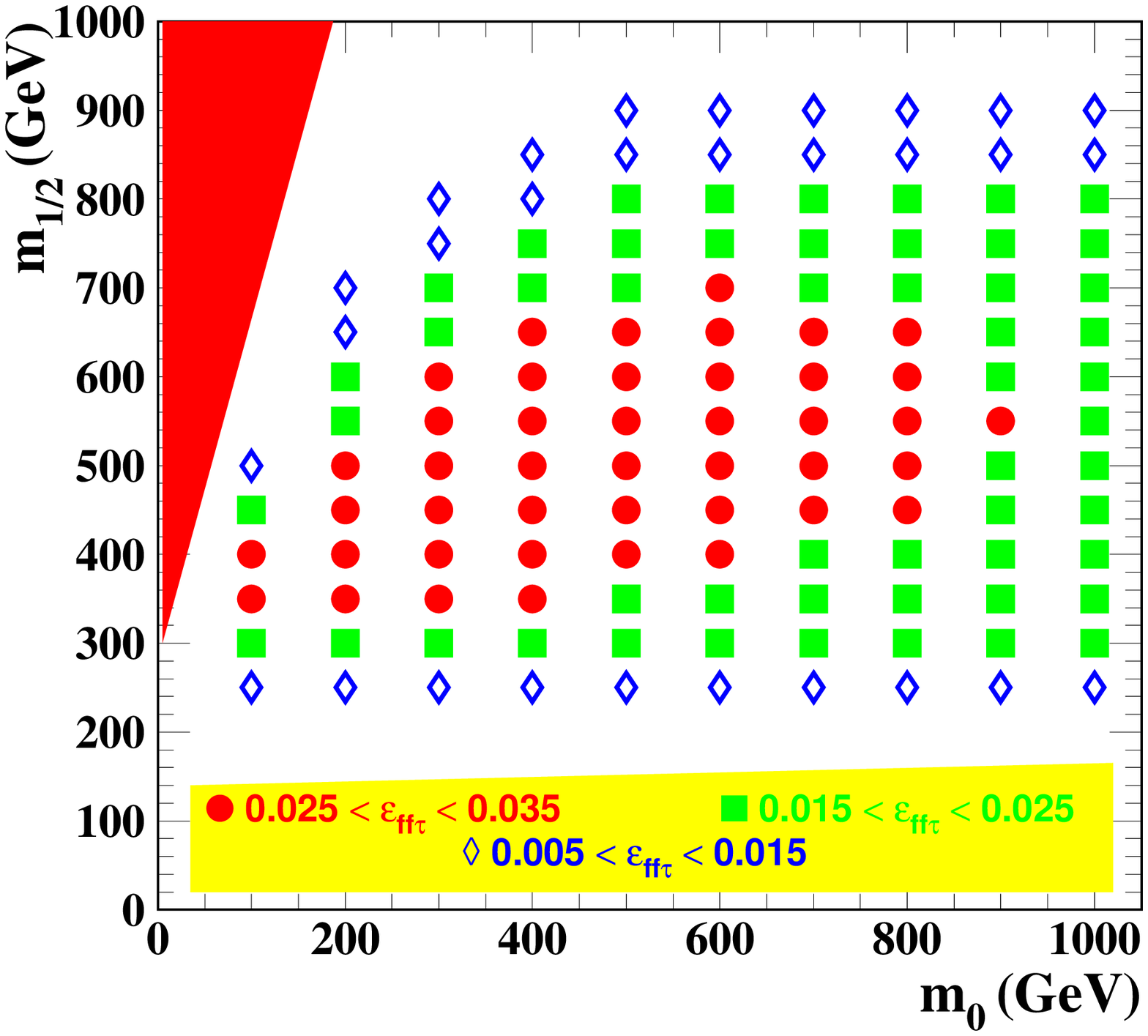}
  \caption{Reconstruction efficiencies of $\mu W$ (left panel) and
    $\tau W$ events (right panel) as a function of $m_0 \otimes
    m_{1/2}$ for $A_0 = -100$ GeV, $\tan\beta=10$ and
    $\mathrm{sgn}(\mu)=+1$ including the effect of pileup. The red
    (dark shaded) area corresponds to the region where stau is the
  LSP, while the yellow (light
  shaded) area represents the region excluded by LEP.}
  \label{fig:eff}
\end{figure}

We evaluated the reconstruction efficiencies as a function of $m_0
\otimes m_{1/2}$ for $A_0 = -100$ GeV, $\tan\beta=10$ and
$\mathrm{sgn}(\mu)=+1$ and our results are depicted in
\fig{fig:eff}. As we can see from the left panel of this
figure, the $\mu W$ decays exhibit a high reconstruction efficiency,
{\em i.e.},  between 10\% and 20\%, in a large area of the parameter
space, degrading only at large $m_{1/2}$. On the other hand, the $\tau
W$ reconstruction (see right panel of \fig{fig:eff}) is at most
3.5\%, indicating that the statistical error on the ratio $R$ is going
to be dominated by these events.

\begin{figure}
  \centering
  \includegraphics[height=10cm,width=80mm]{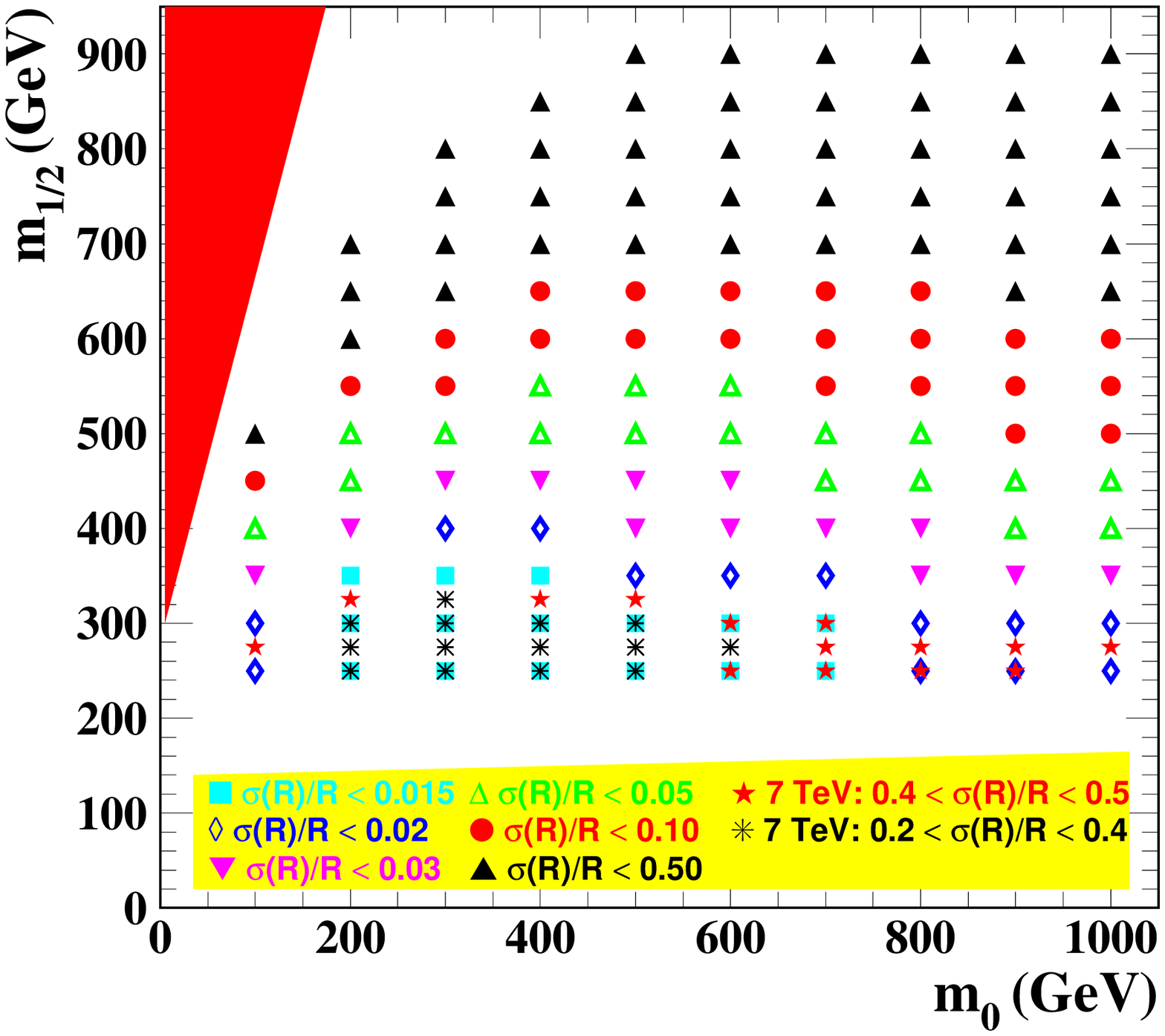}
  \includegraphics[height=10cm,width=80mm]{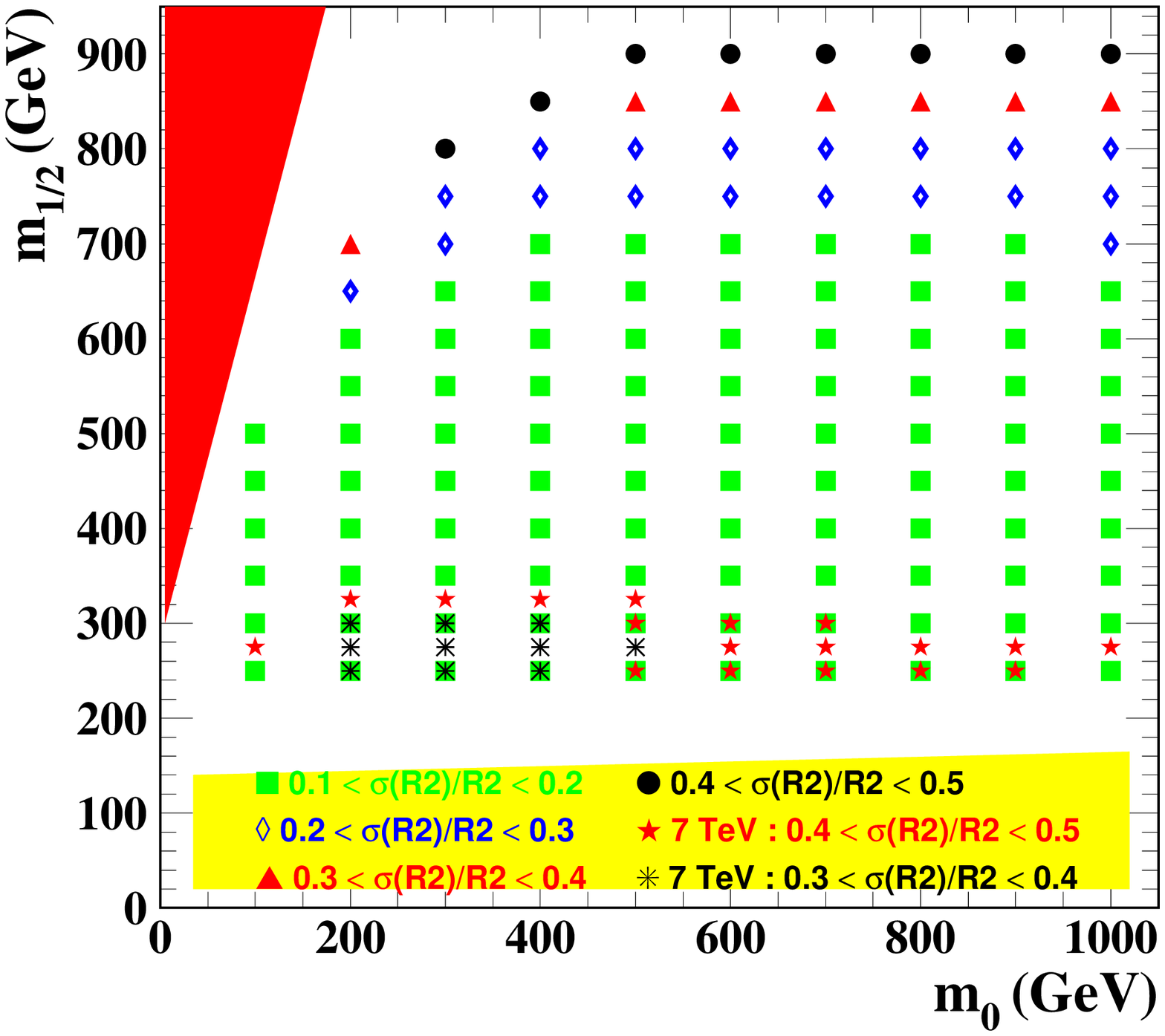}
  \caption{Precision in the determination of the ratio $R$ in the
    plane $m_{1/2}\times m_0$ for a luminosity of $100$ fb$^{-1}$,
    center-of-mass energy of 14~TeV, $A_0 = -100$ GeV, $\tan\beta=10$,
    and $\mathrm{sgn}(\mu)=+1$.  In the right (left) panel we did
    (not) include a possible systematic uncertainty in the extraction
    of the efficiencies for the channels $\mu W$ and $\tau W$. The stars
     in the right panel represent the results for the 7 TeV run with an
    integrated luminosity of 1 fb$^{-1}$. The shaded areas represent
    the same as in \fig{fig:eff}.}
  \label{fig:m12xm0}
\end{figure}

We present in \fig{fig:m12xm0} the attainable precision $\sigma(R)/R$
with which the correlation $R$ can be measured as a function of $m_0
\otimes m_{1/2}$ for $A_0 = -100$ GeV, $\tan\beta=10$, and
$\mathrm{sgn}(\mu)=+1$ for an integrated luminosity of 100 fb$^{-1}$
and a center-of-mass energy of 14 TeV.  We require in all plots that
at least 5 events of reconstructed taus are observed.  In the left
panel of this figure we present the expected statistical error on the
ratio $R$ assuming no systematic errors on the determination of the
reconstruction efficiencies, while in the right panel we consider a
more conservative scenario, where we anticipate a systematic error of
10\% in each of the reconstruction efficiencies.  One can see from
this panel that the precision drops as $m_{1/2}$ grows since the
neutralino production rates from squark/gluino cascade decays also
decrease with increasing $m_{1/2}$ values. Therefore, if the
systematic errors of the efficiency determination are negligible the
LHC collaborations should be able to probe with a very good precision
($\lesssim$ 10\%) the ratio $R$ for $m_{1/2} \lesssim 650$ GeV, which
correspond to an LSP mass up to $\simeq 270$ GeV. The inclusion of systematic
errors at the level assumed in the right panel of \fig{fig:m12xm0}
increases the uncertainty in $R$; however, it is still possible to
perform an accurate test of the RmSUGRA scenario.  

Note that in \fig{fig:m12xm0} we also present results for the 7 TeV run of
the LHC. For this case one can see that the LHC has a much more limited
capability of probing the ratio $R$, since the reach of this run
covers only up to $m_{1/2} \lesssim 300$ GeV. Still, although large,
the statistical errors in this region [$0.3 \lesssim \sigma(R)/R
\lesssim 0.5$], due mainly to the small anticipated integrated
luminosity, which we have taken to be $1$ fb$^{-1}$, 
allow a determination of the atmospheric angle comparable
to that obtained at low energies. \medskip

%

\begin{figure}
  \centering
  \includegraphics[height=70mm,scale=0.4]{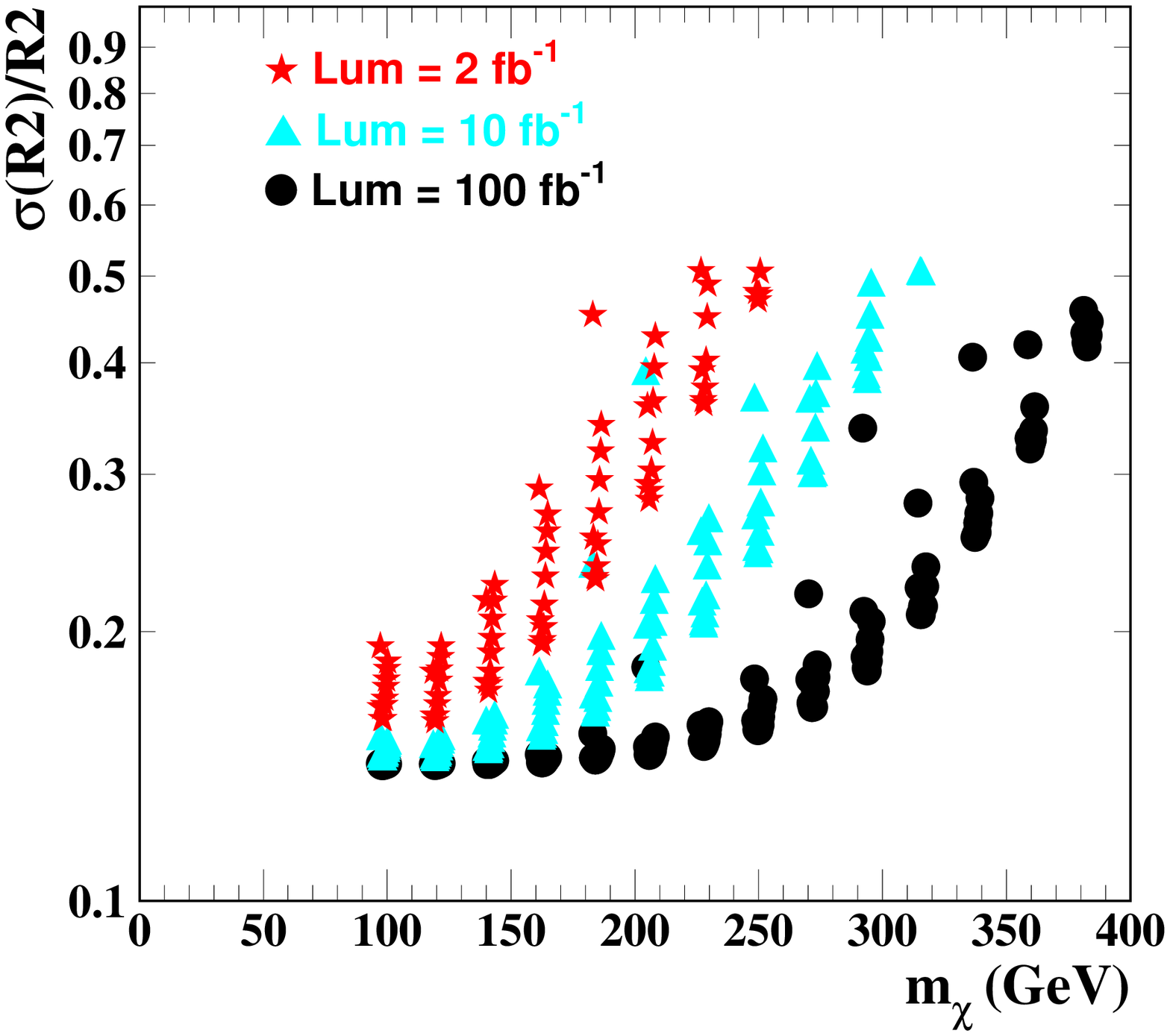}
  \includegraphics[height=70mm,scale=0.4]{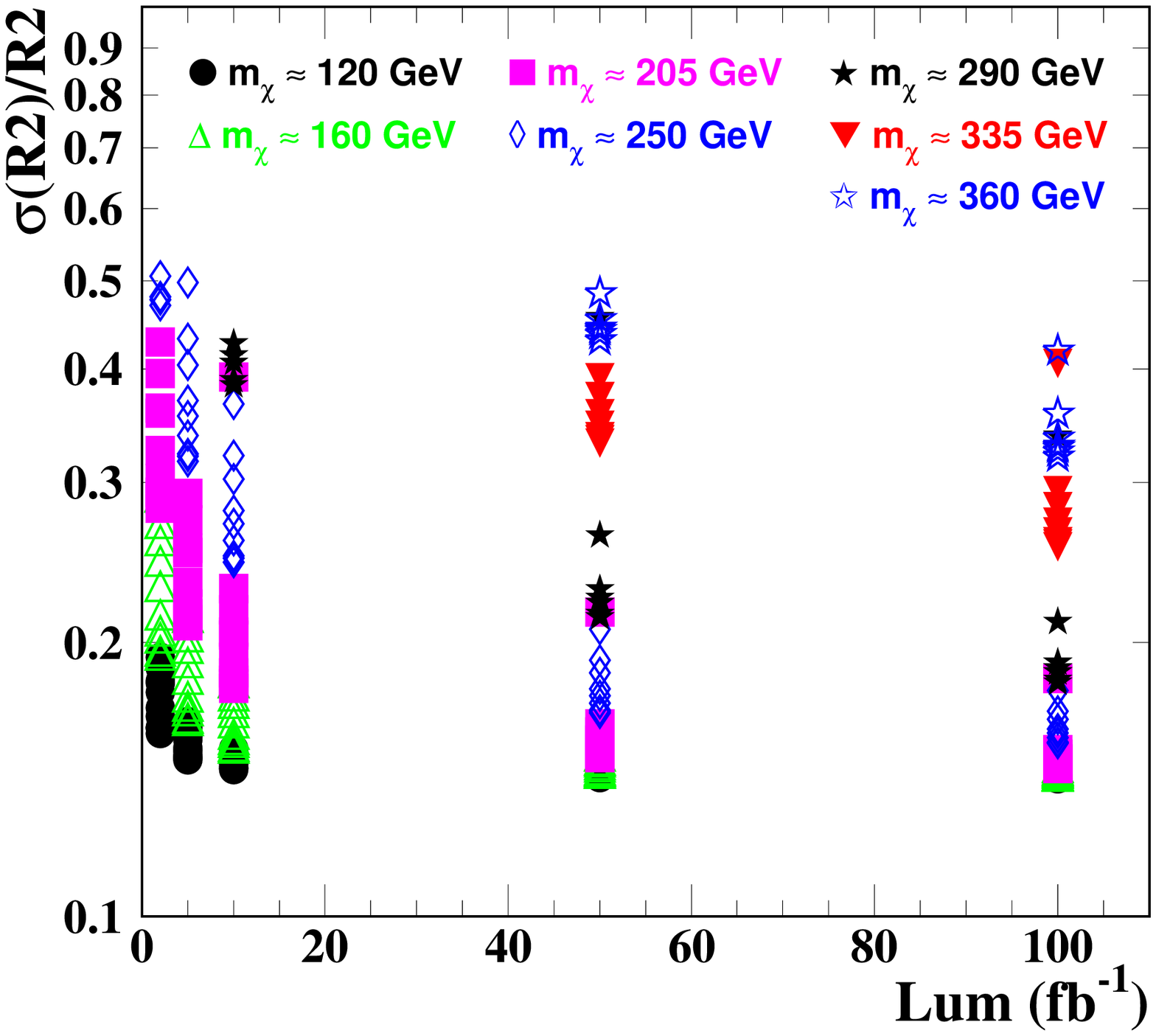}
  \caption{The left panel displays the achievable precision in the
    ratio $R$ as a function of the neutralino mass $m_{\tilde \chi^0_1}$ for
    luminosities of $2$, $10$, and $100$ fb$^{-1}$ at 14 TeV whereas
     the right panel contains the foreseen statistical error
    on $R$ as a function of the integrated luminosity for
    several LSP masses.}
  \label{fig:sigmaxmchi}
\end{figure}

In the left panel of \fig{fig:sigmaxmchi} we show the dependence
of the attained precision as a function of the neutralino mass for
luminosities of $2$, $10$, and $100$ fb$^{-1}$. For small neutralino
masses the SUSY production cross section is large enough to guarantee
that the statistical errors are small; therefore, the uncertainty on
the ratio $R$ is dominated by the assumed systematic errors on the
reconstruction efficiencies, even for an integrated luminosity of 2
fb$^{-1}$.  As the accumulated luminosity increases the LHC
experiments will be able to probe higher neutralino masses;
however, the precision worsens due to the increase of statistical
errors. We can also see clearly that increasing the luminosity allows a
more precise measurement of $R$ as expected. Moreover, one can
probe LSP masses up to 250 (320 or 370) GeV for an integrated
luminosity of 2 (10 or 100) fb$^{-1}$. \medskip

From the right panel of \fig{fig:sigmaxmchi} we estimate the
luminosity  needed to measure $R$ with a given precision for
several LSP masses. For instance, let us consider $m_{\tilde \chi^0_1} = 250$ GeV.
In this case $R$ can only be measured with a precision
$\sigma(R)/R \simeq 50$\% with 2 fb$^{-1}$, while this error
can be brought down to 20\%, {\em i.e.}, close to the limit set by the
systematic uncertainties, with 50 fb$^{-1}$.\medskip

\section{Conclusions}
\label{sec:conc}

We have demonstrated how the LHC may have the potential of probing
neutrino mixing angles with sensitivity competitive to their
low-energy determination by oscillation experiments. This analysis was
carried out, for the sake of concreteness, in the simplest scenario
for neutrino masses induced by minimal supergravity with $R$ parity
violation as framework. In this class of models, the smoking gun for
the neutrino mass generation mechanism is the ratio of branching fractions
of neutralino decaying into $\mu W$ and $\tau W$, as this fraction is
related to the atmospheric neutrino mixing angle in RmSUGRA models.
\medskip

Under realistic detection assumptions we have made a detailed analysis
of the reconstruction of neutralino decays, as well as of the cuts
needed to characterize the signal. After that we determined the
attainable precision on the measurements of the ratio $R$ given in
Eq.~(\ref{eq:corr}). Comparing with a previous parton level study, we
improved the reconstruction efficiencies of muons as well as taus.
\medskip

We showed that the 7 TeV run of the LHC will have a somewhat weak
potential for probing the RmSUGRA model, since it is statistics
limited. Still, precisions comparable to the low-energy determination
should be reached. In contrast, a 14 TeV run with 100 fb$^{-1}$
integrated luminosity will be able to probe a large fraction of the
parameter space with a good precision, as seen in \fig{fig:m12xm0}. In
fact, our analyses suggest that the error on $R$ will be dominated by
the systematic ones on the reconstruction efficiencies of the decay
$\mu W$ and $\tau W$, with the statistical errors being under control.
\medskip

In short, we find that in this case the atmospheric mixing angle may
be probed relatively neatly. In fact, a determination of 
$R$ within a given error translates into a prediction of the 
atmospheric mixing angle with an error of very similar size.
Needless to say, what we have presented is only one example of a
class of LSPs. 
There are other variant schemes based on alternative supersymmetry
and/or $R$ parity breaking, where other states emerge as LSP and similar
correlations to other neutrino mixing angles
appear~\cite{Restrepo:2001me,Hirsch:2002ys, Hirsch:2003fe}. These
would, however, require separate dedicated studies.
%
We encourage the particle detector groups ATLAS and CMS to add the
test of such possibilities to their physics agenda, as this might
lead to a tantalizing synergy between
high-energy~accelerator and low-energy~nonaccelerator searches for
new physics.
Studies with the real LHC data may also make it possible to probe, at
some level, the mass scale characterizing atmospheric neutrino
oscillations, as well as the angle characterizing solar neutrino
oscillations, an issue to be taken up separately.

 \vskip 0.5cm


\noindent\textbf{\large Acknowledgments}\\[.2cm]
\noindent
Work supported in part by Spanish grants FPA2008-00319/FPA, MULTIDARK
Consolider CSD2009-00064 and PROMETEO/2009/091, by European network
UNILHC, PITN-GA-2009-237920, by Conselho Nacional de Desenvolvimento
Cient\'ifico e Tecnol\'ogico (CNPq), and by Funda\c{c}\~ao de Amparo
\`a Pesquisa do Estado de S\~ao Paulo (FAPESP).  F. de Campos thanks
USP and IFIC/C.S.I.C. for hospitality. M.B.M. thanks IFIC/C.S.I.C. for
hospitality. W.P.~is supported by the
DFG, Project No. PO-1337/1-1, and by the Alexander von Humboldt Foundation. 
D.R is partly supported by Sostenibilidad-UdeA/2009 grant.


\end{document}